\documentclass[numberedappendix]{emulateapj} 

\usepackage{hyperref}
\hypersetup{colorlinks,citecolor=Blue,linkcolor=Red,urlcolor=Blue}
\usepackage{amsmath}
\usepackage{empheq}
\usepackage{mathrsfs}
\usepackage[usenames,dvipsnames]{color}

\allowdisplaybreaks 

\begin{document}
 
\title{Simulations of the Solar System's Early Dynamical Evolution with a Self-Gravitating Planetesimal Disk}  
\author{Siteng Fan$^{1}$ \& Konstantin Batygin$^{1}$} 

\affil{$^1$Division of Geological and Planetary Sciences, California Institute of Technology, Pasadena, CA 91125} 

\email{stfan@gps.caltech.edu}

\keywords{planets and satellites: dynamical evolution and stability}

\begin{abstract} 
Over the course of last decade, the Nice model has dramatically changed our view of the solar system's formation and early evolution. Within the context of this model, a transient period of planet-planet scattering is triggered by gravitational interactions between the giant planets and a massive primordial planetesimal disk, leading to a successful reproduction of the solar system's present-day architecture. In typical realizations of the Nice model, self-gravity of the planetesimal disk is routinely neglected, as it poses a computational bottleneck to the calculations. Recent analyses have shown, however, that a self-gravitating disk can exhibit behavior that is dynamically distinct, and this disparity may have significant implications for the solar system's evolutionary path. In this work, we explore this discrepancy utilizing a large suite of Nice odel simulations with and without a self-gravitating planetesimal disk, taking advantage of the inherently parallel nature of graphic processing units. Our simulations demonstrate that self-consistent modeling of particle interactions does not lead to significantly different final planetary orbits from those obtained within conventional simulations. Moreover, self-gravitating calculations show similar planetesimal evolution to non-self-gravitating numerical experiments after dynamical instability is triggered, suggesting that the orbital clustering observed in the distant Kuiper belt is unlikely to have a self-gravitational origin.
\end{abstract} 

\maketitle

\section{Introduction} \label{sec:Introduction}

The narrative of the solar system's formation and dynamical evolution has changed dramatically with the development of the Nice model \citep{Tsiganis2005}. Instead of hitherto conventional in-situ conglomeration theory \citep{Cameron1988} or smooth orbital transport models \citep{Malhotra1993, Hahn2005}, the Nice model proposed a fundamentally violent planetary migration scenario, which entails a series of readily testable observational consequences \citep{Morbidelli2008}. Within the framework of this model, the solar system formed in a more compact configuration with all planetary orbits residing within $\sim$15 au from the sun, encircled by a planetesimal disk that extended to $\sim$30 au. Driven by the scattering of planetesimals, the four outer planets deviated from their initial orbital state, and eventually entered a transient period of dynamically unstable evolution. During this epoch, most of the disk mass was ejected from the solar system as Uranus and Neptune migrated to their current orbits \citep{Levison2008, Levison2011}.

The Nice model represents an important milestone for our understanding of the solar system's formation, as it quantitatively explains the current architecture of the Kuiper Belt \citep{Levison2008, Batygin2011a}, and simultaneously reproduces the non-zero eccentricities and inclinations of the giant planets \citep{Tsiganis2005, Morbidelli2009}. Furthermore, the Nice model successfully accounts for the existence of Jupiter's and Neptune's Trojan asteroids \citep{Morbidelli2005, Nesvorny2007}, and can naturally act as the trigger mechanism of the lunar late heavy bombardment (LHB; \citealt{Gomes2005}). Accordingly, more than a decade after its inception, numerical realizations of the Nice model are plentiful in the literature \citep{Morbidelli2007, Morbidelli2009, Levison2008, Batygin2011a, Batygin2011b, Batygin2012, Nesvorny2012}. In all of the aforementioned studies, self-gravity among planetesimals is neglected, due to its overwhelming computational cost within conventional N-body simulations. Despite this complication, however, disk self-gravity may potentially have important consequences for the evolution of planet orbits and has been a subject of some debate.

\begin{figure*}
\includegraphics[width=\textwidth]{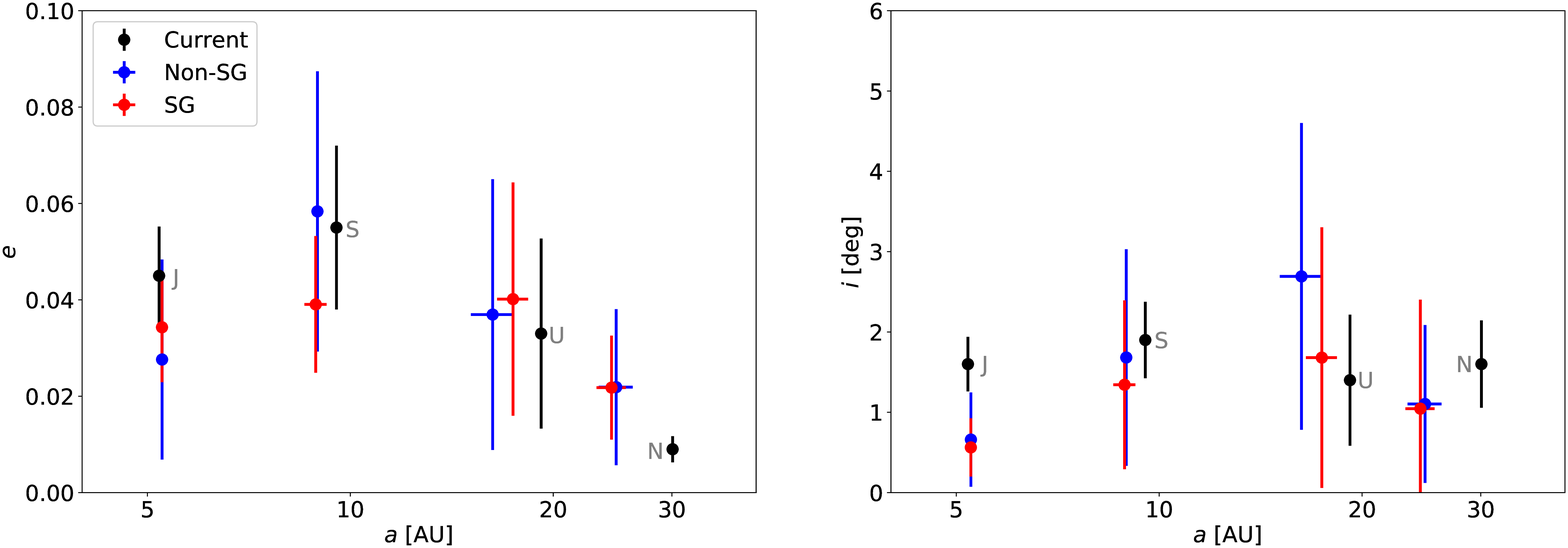}
\caption{Comparison of the final architectures of simulated planetary systems and the current solar system. Blue and red points with error bars represent the orbital elements for non-self-gravitating and self-gravitating cases respectively, computed from the cases which resemble the current solar system. The quantities corresponding to the current solar system are shown by black points with error bars, denoting the amplitudes of secular oscillations of eccentricities and inclinations.}
\label{fig:aei}
\end{figure*}

\citet{Levison2011} carried out the first self-gravitating simulations of the Nice model, employing a simplified algorithm that only considered close encounters among small bodies. Intriguingly, this study demonstrated that intra-particle interactions result in an irreversible exchange of energy between planets and the disk, yielding a natural process for the ignition of the orbital instability. \citet{Reyes-Ruiz2015} performed the first suite of Nice model simulations that treated planetesimal self-gravity in a fully self-consistent manner, and found that upon initiation of the transient instability, one of the two ice giants is consistently ejected from the system, raising concerns regarding the compatibility of a fully self-gravitating Nice model with the real solar system. Finally, the recent study of \citet{Madigan2016} showed that a self-gravitating disk of eccentric planetesimals may be subject to the so-called inclination instability, and can exhibit coupled eccentricity and inclination oscillations reminiscent of global Kozai-Lidov cycles.

Early dynamical evolution aside, the question of whether or not the inclination instability can self-consistently unfold within the solar system is keenly relevant to the present-day observational census of long-period Kuiper belt objects. In particular, the grouping of arguments of perihelion that accompanies the aforementioned eccentricity-inclination cycles of particle orbits has been invoked to explain the observed clustering of arguments among a$>$150 au KBOs \citep{Trujillo2014}, as an alternative to the Planet Nine hypothesis\footnote{In contrast with isolated clustering of the argument of perihelion, the orbital architecture produced by Planet Nine in the distant Kuiper belt is characterized by the simultaneous clustering of the longitudes of perihelion and ascending node (see \citealt{Batygin2017} for a detailed discussion).} \citep{Batygin2016}. To this end, we note that according to \citet{Madigan2016}, the onset of the inclination instability only requires an axisymmetric disk comprising $\sim$1-10$M_\earth$ to be composed of planetesimals on nearly parabolic orbits - a configuration that arises naturally during the early stages of the Nice model instability. Therefore, if the inclination instability can indeed operate in the solar system, it should be captured within the framework of fully self-gravitating Nice model simulations. 

In light of the studies quoted above, the question regarding the role played by the planetesimal disk's self-gravity in the evolution of solar system remains of considerable interest, and answering this question in a statistically significant manner with numerical simulations is the primary purpose of our study. The remainder of the paper is structured as follows. We briefly describe our numerical models in section \ref{sec:Methodology} and present the results of our simulations in section \ref{sec:Results}. We discuss the implications of our calculations in section \ref{sec:Discussion and summary}.

\section{Methodology} \label{sec:Methodology}

To investigate the influence of introducing planetesimal self-gravity into the Nice model, we compare two suites of simulations i.e., one with a self-gravitating planetesimal disk and one with a non-self-gravitating disk. Due to the fundamentally $N^2$ nature of N-body simulations, conventional CPU-based N-body codes are not well suited for fully self-gravitating numerical experiments. However, owing to recent development of the CUDA parallel computing platform and the accompanying graphics processing units \citep{Portegies Zwart2007, Belleman2008}, simulations of the Nice model with fully-self-gravitating primordial planetesimal disks comprised of N$\sim$1000 massive bodies can now be finished within an acceptable amount of time.

To carry out the fully self-gravitating simulations, we utilized the QYMSYM gravitational dynamics software package \citep{Moore2011}. Away from close-approaches, the QYMSYM integrator employs the symplectic Wisdom-Holman mapping \citep{Wisdom1992}, while encounters are handled with a fourth-order Hermite predictor-corrector algorithm \citep{Makino1992}. The timestep was chosen to be $1/(100\pi)$ of Jupiter's orbital period\footnote{We note that such a short timestep is considerable overkill for the simulations at hand. This choice arose due to an error and we have checked that simulations with longer timesteps produce statistically indistinguishable results.} and the calculations were run until orbital equilibration ensued (typically tens of million years). 

The planets were initialized on coplanar ({\it{i}}=0) and near-circular ({\it{e}}$\leq$0.05) orbits within 13 au. In particular, motivated by the results of global hydrodynamical simulations of planet-disk interactions within the solar system \citep{Morbidelli2007}, we adopted a multi-resonant initial condition, where J:S, S:U and U:N period ratios are 2:1, 4:3 and 4:3 respectively. As demonstrated in \citet{Batygin2011a}, successful simulations employing this initial condition lead to a favorable reproduction of the Kuiper belt's dynamical architecture. The massive planetesimal swarm was initialized with 1000 equal-mass particles, comprising a total mass of 30 $M_\earth$ \citep{Tsiganis2005}. The particles were placed on near-coplanar and near-circular orbits with eccentricity ({\it{e}}) and inclination ({\it{i}}) dispersion $\sigma_e$$\sim$$\sigma_i$$\sim$0.01, spanning a radial range from 14 au (near the immediate stability boundary) to 30 au with a surface density inversely proportional to the distance from the sun.

\begin{figure*}
\includegraphics[width=\textwidth]{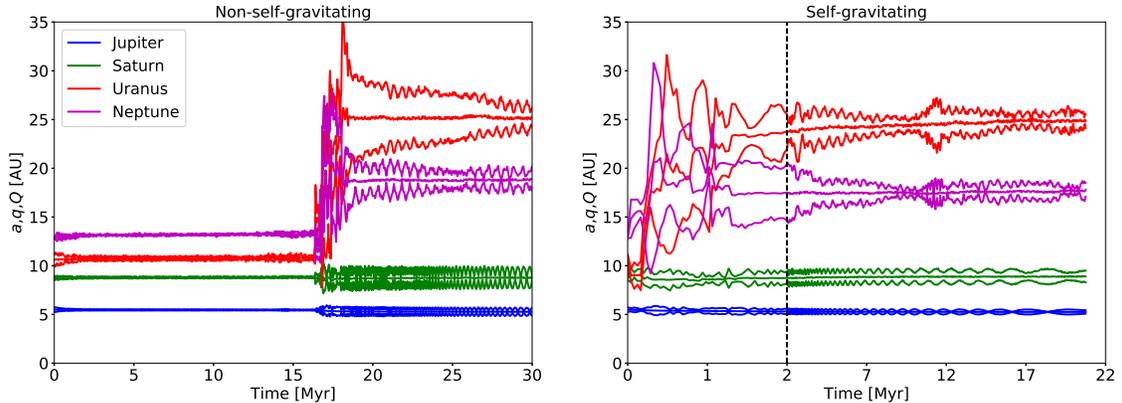}
\caption{Orbital evolution of the solar system with a non-self-gravitating (left) and a self-gravitating (right) primordial planetesimal disk, originating from the same set of initial conditions. Solid lines represent the evolution of semi-major axis ({\it{a}}), as well as perihelion ({\it{q}}) and aphelion ({\it{Q}}) distances of the four gas giants. The x-axis is expanded linearly before 2Myr (left to the dash line) in the right panel in order to more clearly elucidate the planetary evolution at the beginning of the simulation.}
\label{fig:aqQ}
\end{figure*}

The non-self-gravitating simulations were performed using the \textit{mercury6} integrator \citep{Chambers1999}. While this code does not benefit from GPU-based acceleration, it employs essentially the same hybrid symplectic/conventional computational scheme as QYMSYM. One minor difference between the two codes lies in the fact that \textit{mercury6} utilizes the Bulisch-Stoer, rather than the Hermite, algorithm to resolve close encounters. With the exception of this distinction, the two suites of simulations adopted the same exact parameters and initial conditions.

\section{Results} \label{sec:Results}

Having run hundreds of self-gravitating simulations of the Nice model with disks generated each time by drawing {\it{e}} and {\it{i}} randomly from Rayleigh distribution, we identified 48 cases (a little less than 10\% of the total number of simulations) where all four giant planets remained bound to the sun after the post-instability equilibration. Within this set of calculations, 10 runs yielded solar-system-like architectures, wherein all four planets possessed orbital eccentricities smaller than 0.1, inclination less than 6$^{\circ}$  and mean motion ratios of each pair of planets within $\pm$15\% of their true values (Figure \ref{fig:aei}). For comparison, we recomputed the same 48 favorable cases (adopting exactly the same initial conditions), in the non-self-gravitating regime, and identified eight runs that satisfied the above criteria. We note that the success ratio of self-gravitating simulations of the Nice model ($\sim$2\%) is much smaller than that of non-self-gravitating ones, which is 17\% in this work -- comparable to 23\% as reported in \citet{Batygin2010}.

Within this ensemble of 18 integrations, a single initial condition successfully reproduced the outer solar system in both the self-gravitating and non-self-gravitating regimes. The evolutionary tracks of these runs are shown in Figure \ref{fig:aqQ}, where the semi-major axis, as well as perihelion and aphelion distances are plotted for the giant planets as functions of time. Qualitatively, both simulations follow the usual narrative of the Nice model: initialized in a compact configuration, the planets become temporarily unstable, and following a period of chaotic scattering, the orbits circularize due to dynamical friction. 

This sequence of events is representative of all successful runs within our simulation suite. In fact, the final orbital architectures generated in the self-gravitating and non-self-gravitating regimes are essentially indistinguishable from one another (Figure \ref{fig:aei}). Instead, the only clear difference between the two subsets of calculations lies in the time at which the instability is triggered. Particularly, there exists no significant delay between the start of the integration and the onset of planet-planet scattering in the self-gravitating simulation. This discrepancy in the instability time is ubiquitous across all simulations: all runs with non-self-gravitating disks experience a relatively quiescent period of time of $\sim$5-20Myr before planet-planet scattering is initiated. On the contrary, in all calculations with self-gravitating disks, instability is triggered within $\sim$0.1Myr of the starting time. 

Although statistically significant in our calculations, we do not believe that this disparity in instability initiation timescales is physically meaningful. In particular, because we limit our resolution of the planetesimal disk to N=1000 equal-mass bodies, each particle in our simulations is considerably more massive than any real Kuiper belt object. In absence of dynamical friction (that would otherwise ensue in pronounced presence of smaller bodies), the disk self-stirs at an accelerated rate, facilitating the onset of the dynamical instability. As a consequence of this numerical limitation, the reported timescales are not indicative of real instability initiation times and are not meant to coincide with the real timing of the LHB \citep{Gomes2005}. 

\begin{figure}
\includegraphics[width=\columnwidth]{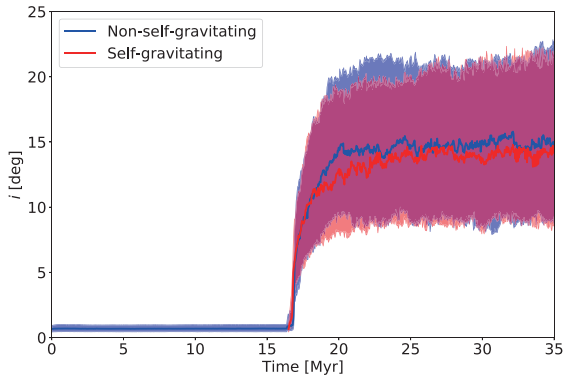}
\caption{Inclination evolution of non-self-gravitating (blue) and self-gravitating (red) primordial planetesimal disks. Solid lines in the center represent the median value of orbital inclination of the planetesimals within the disk. Shadow areas cover the value between the first (25\%) and third (75\%) quartile of orbital inclination of the particles in each case. The plot of the self-gravitating case is offset by 16.5Myr such that both cases have the same apparent instability trigger time.}
\label{fig:ti}
\end{figure}

As already discussed in the Introduction, an intriguing consequence of the dynamical evolution entailed by the Nice model is that through outward scattering, the onset of planet-planet scattering generates an axisymmetric disk of eccentric planetesimals, whose total mass initially comprises $\sim$30$M_\earth$. In essence, this configuration is equivalent to the initial conditions of the inclination instability, considered by \citet{Madigan2016}. Accordingly, we have examined our self-gravitating simulation suite with an eye toward identifying signs of the inclination instability in the calculations. Upon a detailed analysis of particle orbits, we find no indication of the oscillatory behavior of the inclination of planetesimals described by \citet{Madigan2016}. Instead, we find that the evolution of the inclination dispersions of the self-gravitating and non-self-gravitating disks are essentially identical (Figure \ref{fig:ti}). 

\begin{figure}
\includegraphics[width=\columnwidth]{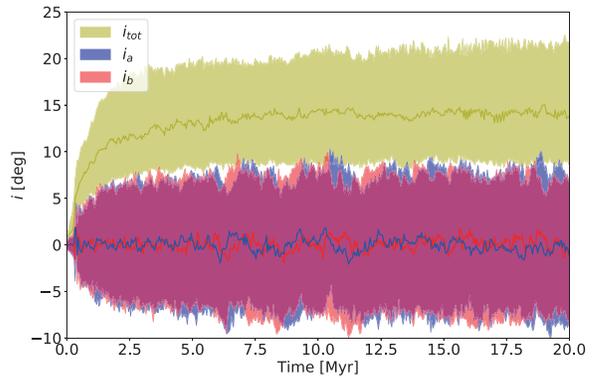}
\caption{Orbital inclination (yellow) and two inclination angles (blue and red) evolution of self-gravitating primordial planetesimal disks. Solid lines in the center represent the median value of orbital inclination or inclination angles of the planetesimals in the disk. Shadow areas cover the first (25\%) and third (75\%) quartiles of the corresponding values.}
\label{fig:Iab}
\end{figure}

In Figure \ref{fig:Iab}, we also present the inclination evolution of the simulation with self-gravitating disk the same way as that in \citet{Madigan2016} via non-standard inclination angles $i_a$ and $i_b$, which are defined as the angles between the ecliptic z-axis and the semi-major/minor axes of the particle orbit. These inclination angles show essentially random circulation after the onset of the Nice model instability, in stark contrast with the coherent behavior observed within the simulations of \citet{Madigan2016}. The fact that the inclination instability fails to operate within the framework of the Nice model can be attributed to two dynamical effects. First, the effective quadrupolar gravitational field generated by the giant planets induces a comparatively rapid precession of the particle orbits, which in turn prevents coherent secular exchange of angular momentum within the disk. Second, there exists a discrepancy in timescales: while the conic structure of a $\sim$10$M_\earth$ disk requires hundreds of million years to develop through secular interactions, close encounters with the giant planets deplete the planetesimal swarm through ejections on a much shorter timescale, generating the low-mass Kuiper belt we observe today.

\section{Discussion and summary} \label{sec:Discussion and summary}

In this work, we have carried out a suite of fully self-gravitating and non-self-gravitating realizations of the Nice model, and have presented a statistically meaningful account of the planetesimal-planetesimal coupling's role in the solar system's early evolution. With lower success rate in self-gravitating simulations, our calculations yield two important insights into the dynamical narrative foretold by the Nice model. First, we find that the inclusion of self-gravity in the numerical experiments yields final solar system configurations that are indistinguishable from those produced within the context of the more conventional, non-self-gravitating simulations. Second, the inclination instability that could potentially arise if the phantasmal disk were to evolve in isolation, is quenched in our simulations due to both secular, as well as short-periodic interactions between the particles and the giant planets.

Although the commencement of planet-planet scattering within the Nice model is envisioned to coincide with the onset of the LHB \citep{Gomes2005}, here we have made no attempt to faithfully reproduce the quiescent period of metastable dynamical evolution that precedes the large-scale dynamical excitation. Nevertheless, the insensitivity of the transient instability's outcome to the details of intra-particle interactions observed in our simulations largely alleviates the concerns brought forth by the simulation suite of \citet{Reyes-Ruiz2015} regarding the compatibility of a fully self-gravitating Nice model with the known structure of the solar system. 

Finally, the failure of the inclination instability to naturally manifest within the framework of the Nice model suggests that the peculiar structure of the distant Kuiper belt is unlikely to have a self-gravitational origin \citep{Madigan2016}. Rather, the existence of Planet Nine appears to be required for the theoretical reproduction of the observational dataset \citep{Batygin2017}. To this end, we note that a scenario wherein the yet-unconfirmed Planet Nine originated at $\sim$10-20 au as the solar system's fifth giant planet \citep{Nesvorny2011, Batygin2012} and was subsequently scattered outward during the Nice model's period orbital rearrangement, remains a distinct possibility \citep{Li2016}. Continued observational unveiling of the distant solar system's dynamical architecture is sure to generate additional constraints that will further inform the feasibility of this sequence of events, and bring the dramatic evolutionary narrative of the solar system into sharper focus. 

\medskip 

\textbf{Acknowledgments:} We thank Christopher Spalding, Michael E. Brown, and Ann-Marie Madigan for useful discussions. Additionally, we would like to thank the anonymous referee for providing a thorough and insightful report that has led to a considerable improvement of the manuscript, as well as the David and Lucile Packard Foundation for their generous support.

\end{document}